\documentclass[12pt]{article}
\usepackage{amssymb}
\usepackage{graphicx}
\usepackage[cp1251]{inputenc}
\usepackage{authblk}

\usepackage{cite}

\newcommand{\be}{\begin{equation}}
\newcommand{\ee}{\end{equation}}
\newcommand{\bea}{\begin{eqnarray}}
\newcommand{\eea}{\end{eqnarray}}
\newcommand{\non}{\nonumber\\}

\title{Electrocaloric effect in KH$_2$PO$_4$}

\author[1]{A.S.Vdovych}
\author[1]{A.P.Moina}
\author[1]{R.R.Levitskii}
\author[2]{I.R.Zachek}
\affil[1]{\textit{Institute for Condensed Matter Physics\\79011, 1
Svientsitskii St, Lviv, Ukraine }} \affil[2]{\textit{Lviv National
Polytechnic University, 12 Bandery Street, 79013, Lviv, Ukraine }}

\date{}

\begin{document}

\maketitle

\sloppy
\begin{abstract}
The proton ordering model for the KH$_{2}$PO$_{4}$ type ferroelectrics is modified by taking into account non-linear effects, namely, the dependence of the effective dipole moments on the proton ordering parameter. Within the four-particle cluster approximation we calculate the crystal polarization, longitudinal dielectric permittivity, specific heat, and explore the electrocaloric effect. Smearing of the ferroelectric phase transition by the longitudinal electric field is described. A good agreement with experiment is obtained.
\end{abstract}



\section{Introduction}

At the moment, the largest electrocaloric (EC) effect, which is the change of  temperature of a dielectric at an adiabatic change of the applied electric field, is observed in perovskite ferroelectrics. Thus, in \cite{Mischenko2006} in the PbZr$_{0.95}$Ti$_{0.05}$O$_{3}$ thin film with
a thickness of 350~nm in a strong electric field (480~kV/cm) the obtained electrocaloric temperature change is $\Delta T=12$~K. \emph{Ab initio} molecular dynamics calculations \cite{Rose2012} predict $\Delta T \thickapprox 20$~K in  LiNbO$_3$. In cheaper and more readily available hydrogen bonded ferroelectrics of the KH$_{2}$PO$_{4}$ (KDP) type the electrocaloric effect was studied for relatively low fields only. Thus, it has been obtained that $\Delta T \thickapprox0.04$~K at $E\thickapprox 4$~kV/cm   \cite{363x}, $\Delta T \thickapprox~1$~K at $E\thickapprox 12$~kV/cm \cite{Baumgartner1950}, and $\Delta T \thickapprox0.25$~K at $T_c$ and $E\thickapprox 1.2$~kV/cm  \cite{Shimshoni1969}. The electrocaloric effect in KDP in high
 fields remains unexplored.

Theoretically the electrocaloric effect in KDP has been described in  \cite{Dunne2008} within the Slater model \cite{Slater1941} and in
 the paraelectric phase only. However, the Slater model is known to give incorrect results in the ferroelectric phase. Influence of electric field
on the thermodynamic characteristics of the KDP type crystals, such as polarization, dielectric permittivity, piezoelectric coefficients, elastic constants
 has been described in \cite{Stasyuk2001,0311U6,JPS1701} within the proton ordering model with the piezoelectric coupling to the shear strain
  $\varepsilon_6$ and proton tunneling \cite{lis2007} taken into account. However, these theories required, in particular, invoking two different values
  of   the effective dipole moments for the paraelectric and ferroelectric phase \cite{Stasyuk2001,JPS1701}. This made a correct description of the system
  behavior in the fields high enough to smear out the first order phase transition impossible.

In the present paper we suggest a way to circumvent this difficulty. Assuming that the difference between the dipole moments is caused by non-zero
values of the order parameter, we modify the proton ordering model accordingly. The crystal characteristics in zero field and in high fields are calculated.
 Smearing of the first order phase transition and the electrocaloric effect are described.

\section{Thermodynamic characteristics}

We consider the KDP type ferroelectrics in presence of the external shear stress
 $\sigma_6 = \sigma_{xy}$ and electric field $E_3$ applied along the crystallographic axis
 $\textbf{c}$, inducing the strain
 $\varepsilon_6$ and polarization $P_3$.
The total model Hamiltonian reads \cite{0311U6}
 %% 2.1
% \setcounter{equation}{0}
% \renewcommand{\theequation}{2.\arabic{equation}}
 \be
 \hat H = N\hat H_{0} + \hat H_s,
 \ee
 where $N$ is the total number of primitive cells;  The ``seed''
energy corresponds to the sublattice of heavy ions and does not depend explicitly
on the deuteron subsystem configuration. It is expressed in terms of the strain $\varepsilon_6$ and electric field
 $E_3$ and includes the elastic, piezoelectric, and dielectric contributions
 %% 2.2
 \be
 \hat H_{0} =  {v} \left( \frac12 c_{66}^{E0}\varepsilon_6^2 -
 e_{36}^0E_3\varepsilon_6 - \frac12 \chi_{33}^{\varepsilon 0}E_3^2
 \right),
 \ee
 where $v$ is the
primitive cell volume; $c_{44}^{E0} $, $e_{36}^0$, ${\chi }_{33}^{\varepsilon 0} $ are the ``seed'' elastic constant,
piezoelectric coefficient, and dielectric susceptibility.

The pseudospin part of the Hamiltonian reads
 %%% 2.3
 \bea
 && \!\!\!\! \!\!\!\! \!\!\!\!\!\!\!\! \hat H_s = \frac12 \sum\limits_{ {qf}\atop{q'f'}}J_{ff'}(qq')
 \frac{\sigma_{qf}}{2}\frac{\sigma_{q'f'}}{2} + \hat H_{sh} + \sum\limits_{qf} 2\psi_6\varepsilon_6 \frac{\sigma_{qf}}{2}
 - \sum\limits_{qf}\mu_{f}E_3 \frac{\sigma_{qf}}{2} + \hat H_E. \label{H_s}
 \eea

Here the first term describes the effective long-range interactions between  protons, including
also indirect lattice-mediated interactions  \cite{122x,133x}, $\sigma_{qf}$ is the operator of the
 $z$-component of a pseudospin, corresponding to the proton on the $f$-th hydrogen bond ($f$=1,2,3,4) in the  $q$-th
cell. Its eigenvalues $\sigma_{qf}= \pm 1$
  are assigned to two equilibrium positions of a proton on this bond

In (\ref{H_s}) $\hat H_{sh}$ is the Hamiltonian of the short-range interactions between
 protons, which includes  linear over the strain
 $\varepsilon_6$ terms  \cite{0311U6,JPS1701}
 \bea
 && \hspace{-4ex} \hat H_{sh} = \sum\limits_q \Bigl\{ \left(
 \frac{\delta_{s}}{8}\varepsilon_6 +
 \frac{\delta_{1}}{4}\varepsilon_6\right) (\sigma_{q1} + \sigma_{q2} + \sigma_{q3} +
 \sigma_{q4}) +\non
 && \hspace{-4ex} +\! \left( \frac{\delta_{s}}{8}\varepsilon_6 - \frac{\delta_{1}}{4}\varepsilon_6\right)
 (\sigma_{q1} \sigma_{q2} \sigma_{q3} \!+\! \sigma_{q1} \sigma_{q2} \sigma_{q4}
 \!+\! \sigma_{q1} \sigma_{q3} \sigma_{q4} \!+\! \sigma_{q2} \sigma_{q3}
 \sigma_{q4}) + \non
 && \hspace{-4ex} + \frac14 (V + \delta_{a}\varepsilon_6)(\sigma_{q1} \sigma_{q2} + \sigma_{q3} \sigma_{q4}
 )+ \frac14 (V - \delta_{a}\varepsilon_6)(\sigma_{q2} \sigma_{q3} + \sigma_{q4}
 \sigma_{q1}) + \\
 && \hspace{-4ex} +\! \frac14 U (\sigma_{q1} \sigma_{q3} \!+\! \sigma_{q2}
 \sigma_{q4}) \!+\! \frac{1}{16} \Phi \sigma_{q1} \sigma_{q2} \sigma_{q3}
 \sigma_{q4} \Bigr\}.\nonumber
 \eea
Here
 \[
 V = - \frac12 w_1, ~~ U = \frac12 w_1 - \varepsilon, ~~ \Phi =
 4\varepsilon- 8w + 2w_1,
 \]
where $\varepsilon$ $w$, $w_1$ are the energies of proton configurations.

The third term in (\ref{H_s}) is a linear over the shear strain
$\varepsilon_6$ field due to the piezoelectric coupling; $\psi_6$
is the deformational potential \cite{0311U6}.

The fourth term in  (\ref{H_s}) effectively describes the system interaction with the external electric field $E_3$. Here $\mu_{f}$
is the effective dipole moment of the  $f$-the hydrogen bond, and
 \[
 \mu_{1} = \mu_{2} = \mu_{3} = \mu_{4} = \mu.
 \]

The fifth term in (\ref{H_s}) is introduced in the present paper for the first time. It takes into account the dependence of the
effective dipole moment on the order parameter (pseudospin mean value)
\be \hat H_E =- \frac{1}{N^2}\sum\limits_{qf} \left(\sum\limits_{q'f'}\frac{\sigma_{q'f'}}{2}\right)^2 \mu'E_3 \frac{\sigma_{qf}}{2}. \label{H_E}\ee

Considering the crystal structure of the KDP type ferroelectric,
the four-particle cluster approximation is most suitable for the
short-range interactions \cite{133x,46x}. The long-range
interactions and the term $\hat H_E$ are taken into account in the
mean field approximation. Thus,
 \bea
 && \hat H_E=- \frac{1}{N^2}\sum\limits_{qf} \left(\sum\limits_{q'f'}\frac{\sigma_{q'f'}}{2}\right)^2 \mu'E_3 \frac{\sigma_{qf}}{2} =  - \frac{1}{N^2}\frac{\mu'E_3}{8} \sum\limits_{qf} \sum\limits_{q'f'} \sum\limits_{q''f''} \sigma_{qf} \sigma_{q'f'} \sigma_{q''f''} \approx \nonumber\\
 && - \frac{1}{N^2}\frac{\mu'E_3}{8} \sum\limits_{qf} \sum\limits_{q'f'} \sum\limits_{q''f''} ((\sigma_{qf}+\sigma_{q'f'}+\sigma_{q''f''})\eta^2 - 2\eta^3) =
\nonumber\\
 && - N\frac{\mu'E_3}{8} \sum\limits_{f=1}^4 \sum\limits_{f'=1}^4 \sum\limits_{f''=1}^4 ((\sigma_{f}+\sigma_{f'}+\sigma_{f''})\eta^2 - 2\eta^3) =
\nonumber\\
 && - 12N\mu'E_3 \sum\limits_{f=1}^4 \frac{\sigma_{qf}}{2} \eta^2
+ 16N\mu'E_3 \eta^3.
 \eea

The calculated thermodynamic potential per one primitive cell reads
 \bea  && G =  H^{(0)}
+ 2\nu_c\eta^2 + 16\mu'E_3 \eta^3 +
\frac{1}{2\beta}\sum\limits_{f=1}^4\ln Z_{1f} - \frac1\beta\ln
Z_4- { {v}}\sigma_6\varepsilon_6, \label{Gz} \eea where $4\nu_c =
J_{11}(0) + 2J_{12}(0) + J_{13}(0)$ is the eigenvalue of the
long-range interactions matrix Fourier transform $J_{ff'} =
\sum\limits_{ {\bf R}_q - {\bf R}_{q'}} J_{ff'}(qq')$;
 \[
 \eta = \langle \sigma_{q1} \rangle = \langle \sigma_{q2} \rangle
 = \langle \sigma_{q3} \rangle = \langle \sigma_{q4} \rangle
 \]
is the proton ordering parameter; $Z_{1f}=Sp e^{-\beta\hat
H_{qf}^{(1)}}$, $Z_{4}=Sp e^{-\beta\hat H_{q}^{(4)}}$ are the
single-particle and four-particle  partition functions;
$\beta=\frac{1}{k_BT}$. The single-particle $\hat H_{qf}^{(1)}$
and four-particle $\hat H_{q6}^{(4)}$ proton Hamitonians are
 \be
 \hat H_{qf}^{(1)} = - \frac{ \bar z_{f}}{\beta}\frac{\sigma_{qf}}{2},
 \ee
 \bea
 && \hspace{-4ex} \hat H_{q}^{(4)} = - \sum\limits_{f=1}^4 \frac{z }{\beta}
 \frac{\sigma_{qf}}{2}+ \frac{\varepsilon_6}{4} (-\delta_{s} +
 2\delta_{1}) \sum\limits_{f=1}^4\frac{\sigma_{qf}}{2} - \\
 && \hspace{-4ex} -\! \varepsilon_6(\delta_{s} \!+\! 2\delta_{1})\!
 \left(
 \frac{\sigma_{q1}}{2}\frac{\sigma_{q2}}{2}\frac{\sigma_{q3}}{2}\!+\!
 \frac{\sigma_{q1}}{2}\frac{\sigma_{q2}}{2}\frac{\sigma_{q4}}{2} \!+\!
 \frac{\sigma_{q1}}{2}\frac{\sigma_{q3}}{2}\frac{\sigma_{q4}}{2} \!+\!
 \frac{\sigma_{q2}}{2}\frac{\sigma_{q3}}{2}\frac{\sigma_{q4}}{2} \right)
 + \non
 && \hspace{-4ex} +\! (V \!+\! \delta_{a}\varepsilon_6)
 \left( \frac{\sigma_{q1}}{2}\frac{\sigma_{q2}}{2} \!+\!
 \frac{\sigma_{q3}}{2}\frac{\sigma_{q4}}{2}\right) \!+\! (V \!-\! \delta_{a}\varepsilon_6)
 \left( \frac{\sigma_{q2}}{2}\frac{\sigma_{q3}}{2} \!+\!
 \frac{\sigma_{q4}}{2}\frac{\sigma_{q1}}{2}\right) \!+\non
 && \hspace{-4ex} + U \left( \frac{\sigma_{q1}}{2}\frac{\sigma_{q3}}{2} +
 \frac{\sigma_{q2}}{2}\frac{\sigma_{q4}}{2}\right) + \Phi
 \frac{\sigma_{q1}}{2}\frac{\sigma_{q2}}{2}\frac{\sigma_{q3}}{2}\frac{\sigma_{q4}}{2},
 \nonumber
 \eea
where
 \[
 z  = \beta [- \Delta^c + 2\nu_c \eta - 2\psi_6\varepsilon_6 +
 \mu E_3 + 12\mu'\eta^2 E_3 ],
 \]
 \[
 \bar z_{f} = \beta [- 2 \Delta^c + 2\nu_c \eta - 2\psi_6\varepsilon_6 + \mu E_3 + 12\mu'\eta^2 E_3 ].
 \]

The effective field $ \Delta^c$ exerted by the neighboring
hydrogen bonds from outside the cluster can be determined from the
self-consistency condition: the pseudospin mean value $\langle
\sigma_{qf} \rangle$ calculated with the four-particle and with
the one-particle Hamiltonians must coincide
 \be
 \langle \sigma_{qf} \rangle = \frac{ {\rm Sp} \left\{ \sigma_{qf}
 e^{ -\beta \hat H_{q}^{(4)}}\right\}} { {\rm Sp}\, e^{-\beta \hat
 H_{q}^{(4)}}} = \frac{ {\rm Sp} \left\{ \sigma_{qf}
 e^{ -\beta \hat H_{qf}^{(1)}}\right\}} { {\rm Sp}\, e^{-\beta \hat
 H_{qf}^{(1)}}}.
 \ee
Finally, the order parameter is
 \be
 \eta = \frac{m }{D},
 \ee
where
 \bea
 && \hspace{-4ex} m  = \sinh (2z  + \beta \delta_{s}\varepsilon_6) + 2b \sinh(z  - \beta
 \delta_{1}\varepsilon_6), \non
 && \hspace{-4ex} D = \cosh (2z  + \beta \delta_{s}\varepsilon_6) + 4b \cosh
 (z  - \beta \delta_{1}\varepsilon_6) + 2a \cosh \beta
 \delta_{a}\varepsilon_6 + d, \non
 && \hspace{-4ex} z  = \frac12 \ln \frac{1 + \eta}{1 - \eta}
 + \beta \nu_c\eta - \beta\psi_6\varepsilon_6 + \frac{\beta
 \mu }{2}E_3  + 6\beta\mu'\eta^2 E_3, \non
 && \hspace{-4ex} a = e^{-\beta \varepsilon}, ~~ b = e^{-\beta w}, ~~ d = e^{-\beta
 w_1}. \nonumber
 \eea

The thermodynamic potential  (\ref{Gz}) is then obtained in the following form
 \bea && G = \frac{ {v}}{2}c^{E0}_{66}\varepsilon^2_6 -
{{v}}e^0_{36}\varepsilon_6E_3
- \frac{{v}}{2}\chi^{\varepsilon 0}_{33}E^2_3 + 2{{\nu}}_c\eta^2 + 16\mu'E_3 \eta^3 + \label{G}\\
&&\quad{}+ \frac2\beta\ln 2 - \frac2\beta\ln[1- \eta^2] -
\frac2\beta\ln D- {{v}}\sigma_6\varepsilon_6. \nonumber \eea From
the condition of the thermodynamic potential minimum
\[
 \left( \frac{\partial
G}{\partial\varepsilon_6} \right)_{T,E_{3},\sigma_6} = 0
\]
we obtain an equation for the strain  $\varepsilon_6$
 \bea
%\label{3.2}
&& \hspace{-8ex} \sigma_6 \!=\! c^{E0}_{66}\varepsilon_6 \!-\!
e^0_{36}E_3 \!+\! \frac{4\psi_6}{v} \eta \!+\!\frac{2r }{ v D}.
\label{s_6}
 \eea
In the same way we derive the expressions for polarization $P_3$
and molar entropy of the proton subsystem
% 3.2  3.3
\bea
%\label{3.2}
&&  \hspace{-8ex}
 P_3 = - \frac{1}{ {v}}\left( \frac{\partial G}{\partial E_3}
\right)_{T,\sigma_6} = e^0_{36}\varepsilon_6 + \chi^{\varepsilon 0}_{33}E_3 +
2\frac{\mu}{v}\eta +
8\frac{\mu'}{v}\eta^3, \label{P_3} \\
&& \hspace{-8ex}  S = - \frac{N_A}2\left( \frac{\partial
G}{\partial T}\right)_{E_3,\sigma_6} = R \left\{ -\ln2 + \ln [1 -
\eta^2] + \ln D +  2Tz_T\eta + \frac{M }{D} \right\}. \label{S}
 \eea
Here $N_A$ is the Avogadro number; $R$ is the gas constant. The
following notations are used
\bea && r =-\delta_{s}M_{s}-\delta_{a}M_{a}+\delta_{1}M_{1}, \nonumber \\
&&z_T = - \frac{1}{k_B T^2} ({\nu}_c \eta -
\psi_6\varepsilon_6 + 6\mu'\eta^2 E_3), \nonumber \\
&&
 M  = 4b\beta w\cosh(z  - \beta
 \delta_{1}\varepsilon_6) + \beta w_1d + 2a\beta\varepsilon\cosh\beta
 \delta_{a}\varepsilon_6+\beta
 \varepsilon_6r , \nonumber \\
&&  M_{a}=2a\sinh \beta\delta_{a}\varepsilon_6, M_{s}=\sinh (2z
+\beta\delta_{s}\varepsilon_6), M_{1}=4b \sinh (z
-\beta\delta_{1}\varepsilon_6). \nonumber \eea

From Eqs.~(\ref{s_6}), (\ref{P_3}) we find the isothermal dielectric susceptibility of a clamped crystal ($\varepsilon_6={\rm
 const}$):
 \be
 \chi^{T\varepsilon}_{33} = \left(\frac{\partial P_3}{\partial
E_3}\right)_{T,\varepsilon_6} = \chi^0_{33} + \frac{(\mu +
12\mu'\eta^2)^2}{v} \frac{2\beta\varkappa }{D- 2\varkappa
z_{\eta}},  \ee where \bea && \varkappa  = \cosh (2z
+\beta\delta_{s}\varepsilon_6)
 + b\cosh (z  -\beta\delta_{1}\varepsilon_6) - \eta m , \nonumber \\
&&  z_{\eta} = \frac{1}{1- \eta ^2} + \beta\nu_c + 12\beta\mu'\eta
E_3;\nonumber \eea the isothermal piezoelectric coefficient
$e_{36}^{T}$
% 3.10
\bea && \hspace{-4ex} e_{36}^T = - \left(
\frac{\partial\sigma_6}{\partial E_3} \right)_{T,\varepsilon_6} =
\left( \frac{\partial P_3}{\partial\varepsilon_6}\right)_{T,E_3} =
e^0_{36} + \frac{2(\mu + 12\mu'\eta^2)}{v}
\frac{\beta\theta_6}{D-2 z_{\eta}\varkappa }. \eea where \bea &&
\theta_6 = - 2 \varkappa \psi_6 + f_6,~~~~ f_6=\delta_{s} \cosh
(2z  +\beta\delta_{s}\varepsilon_6) - 2b\delta_{1}\cosh(z
-\beta\delta_{1}\varepsilon_6)+\eta r ;\nonumber \eea the
isothermal elastic constant at constant field
 \bea
 && c_{66}^{TE} = c_{66}^{E0} + \frac{8\psi_6}{v}
 \frac{\beta(- \psi_6 \varkappa  + f_6)}{D - 2 z_{\eta}
 \varkappa } -
 \frac{4 \beta z_{\eta} f_6^2}{v D (D - 2 z_{\eta}
 \varkappa )} -\\
 && - \frac{2\beta}{v D} [\delta_{s}^2 \cosh (2z  + \beta \delta_{s} \varepsilon_6) +
2a \delta_{a}^2 \cosh \beta\delta_{a}\varepsilon_6 + 4b
\delta_{1}^2 \cosh (z  - \beta \delta_{1} \varepsilon_6)
 ] + \frac{2\beta r ^2}{v D^2}. \nonumber
 \eea

Other isothermal dielectric and piezoelectric characteristics can
be expressed via those found above, using the known thermodynamic relations. Thus, the isothermal dielectric susceptibility
of a free crystal ($\sigma_6$=const)
 \be \chi^{T\sigma}_{33} = \left(\frac{\partial
P_3}{\partial E_3} \right)_{T,\sigma_6} = \chi^{T\varepsilon}_{33}
+ \frac{(e_{36}^T)^2}{c_{66}^{TE}} = \chi^{T\varepsilon}_{33} +
e_{36}^Td_{36}^T, \ee
isothermal piezoelectric coefficient
 \be
d_{36}^T = \left(\frac{\partial \varepsilon_6}{\partial E_3}
\right)_{T,\sigma_6} = \left(\frac{\partial
P_3}{\partial\sigma_6}\right)_{T,E_3} =
\frac{e_{36}^T}{c_{66}^{TE}},
 \ee

%скбм№-Ўсъп- │м-v- и'скб№v№№м--ъпбж п-и-бЎ-
% \be
%h_{36}^T = -\left(\frac{\partial E_3}{\partial \varepsilon_6}
%\right)_{T,P_3} = -\left(\frac{\partial\sigma_6}{\partial
%P_3}\right)_{T,\varepsilon_6} =
%\frac{e_{36}^T}{\chi_{33}^{T\varepsilon}}, \ee
%
%скбм№-Ўсъп- │м-v- и'скб№v№№м--ъпбж а№рб-Ў-мсж
% \be g_{36}^T = -
%\left(\frac{\partial E_3}{\partial\sigma_6}\right)_{T,P_3}
%=\left(\frac{\partial\varepsilon_6}{\partial
%P_3}\right)_{T,\sigma_6} =
%\frac{e_{36}^T}{\chi^{T\varepsilon}_{33}c^{TE}_{66} +
%(e^{T}_{36})^2} = \frac{h_{36}^T}{c^{TP}_{66}},  \ee
%
%скбм№-Ўсъп- и-б┐п- │м-v- и-- иб│мспс ибvш--к-мсж
% \be
% c_{66}^{TP} = \left(\frac{\partial \sigma_6}{\partial \varepsilon_6}
%\right)_{T,P_3} = c_{66}^{TE} + e_{36}^Th_{36}^T = c_{66}^{TE} +
%\frac{(e_{36}^T)^2}{\chi_{33}^{T\varepsilon}},
% \ee
%
%скбм№-Ўсъпс иба-мv-┐б│мс и-- иб│мспбЎб ибvс:
% \be s^{TE}_{66} =
%\left(\frac{\partial\varepsilon_6}{\partial\sigma_6}\right)_{T,E_3}
%= \frac{1}{c^{TE}_{66}}, ~~~~ s^{TP}_{66} =
%\left(\frac{\partial\varepsilon_6}{\partial\sigma_6}\right)_{T,P_3}
%= \frac{1}{c^{TP}_{66}}. \ee

The molar specific heat of the proton subsystem is
%  4.3
 \be \Delta
C^\sigma = T\left( \frac{\partial S}{\partial T}\right)_{\sigma} =
T(S_T + S_{\eta}\eta_T + S_{\varepsilon}\varepsilon_T), \ee
Here we used the following notations
% 4.5
\bea && S_T = \left(\frac{\partial S}{\partial T}
\right)_{P_3,\varepsilon_6} =\frac{R}{DT}\left\{ 2Tz_T (q_6 - \eta
M ) +
N_6 - \frac{M ^2}{D} \right\}, \\
&& S_{\eta} = \left(\frac{\partial S}{\partial \eta}
\right)_{\varepsilon_6,T} = \frac{2R}{D} \{ DTz_T + [q_6 - \eta M
] z_{\eta}\} \nonumber \\
&&
 S_{\varepsilon} \!=\! \left(\frac{\partial S}{\partial\varepsilon_6}\right)_{\eta,T} \!=\!
 \frac{R}{k_B T D} \left\{ - 2[q_6 - \eta M ]
\psi_6\!-\! \lambda \!+\!\frac{M }{D} r  \right\}, \nonumber
\end{eqnarray}
\bea && \hspace{-5ex} N_6
=2a(\beta\varepsilon)^2\cosh\beta\delta_{a}\varepsilon_6 +4
b(\beta w)^2\cosh(z  -\beta\delta_{1}\varepsilon_6) +(\beta w_1)^2d+\nonumber\\
&&
 \hspace{-5ex}
+2\beta^2\varepsilon_6
(-\varepsilon\delta_{a}M_{a}+w\delta_{1}M_{1}) +
\nonumber\\
&& \hspace{-5ex} +\varepsilon_6^2
[2a(\beta\delta_{a})^2\cosh\beta\delta_{a}\varepsilon_6 +
(\beta\delta_{s})^2\cosh(2z  +\beta\delta_{s}\varepsilon_6) + 4
b(\beta\delta_{1})^2\cosh(z  -\beta\delta_{1}\varepsilon_6)],
\nonumber\\
&& \hspace{-5ex} q_6 = 2b\beta w \sinh(z
-\beta\delta_{1}\varepsilon_6)
+\beta\varepsilon_6[-\delta_{s}\cosh(2z
+\beta\delta_{s}\varepsilon_6) + 2b\delta_{1}\cosh(z
-\beta\delta_{1}\varepsilon_6)],
 \nonumber\\
&& \hspace{-5ex}
\lambda =-\beta\varepsilon\delta_{a}M_{a}+\beta
w\delta_{1}M_{1} +\nonumber\\
&& \hspace{-5ex} +\varepsilon_6 \beta[\delta_{s}^2\cosh(2z
+\beta\delta_{s}\varepsilon_6) +
2a\delta_{a}^2\cosh\beta\delta_{a}\varepsilon_6 + 4b\delta_{1}^2
\cosh(z  -\beta\delta_{1}\varepsilon_6)],
 \nonumber\\
&& \hspace{-5ex}
 \eta_T = p_6^\varepsilon + \frac{v}{2(\mu+12\mu'\eta^2)} [e_{36}^T-e_{36}^0]\varepsilon_T,   \nonumber\\
&& \hspace{-5ex}
 \varepsilon_T = \left(\frac{2}{vDT} (2Tz_Tf_6-\lambda +\frac{M r }{D}) - \frac{4 p_6^{\varepsilon}}{v}(\psi_6-\frac{z_{\eta}f_6}{D})\right)/c_{66}^{TE},
 \nonumber\\
&& \hspace{-5ex}
 p_6^{\varepsilon} = \frac1T\frac{2\varkappa Tz_T + [q_6 -\eta M ]}{D-2\varkappa  z_{\eta}}. \eea

The total specific heat is the sum of the proton and lattice contributions
\be C=\Delta C^\sigma + C_{lattice}  \ee The lattice heat capacity
near  $T_c$ is approximated by a linear dependence \be C_{lattice}
= C_0+C_1(T-T_c) \ee
Then the lattice entropy near  $T_c$ is
\be S_{lattice} = \int \frac{C_{lattice}}{T}dT = (C_0-C_1T_c)\ln(T)+C_1T + const \label{Stotal1}\ee
The total entropy is a function of temperature and electric field
\be S_{total}(T,E)=S + S_{lattice} \label{Stotal} \ee
Solving  Eq.(\ref{Stotal}) with respect to temperature at
$S_{total}(T,E)=const$ and two different fields, we can find the
electrocaloric temperature change
\be \Delta T = T(S_{total},E_2)-T(S_{total},E_1). \label{DT_S} \ee

Alternatively, the electrocaloric temperature change can be calculated using the know formula
\be \Delta T = \int \limits_{0}^{E} \frac{TV}{C} \left(\frac{\partial P_3}{\partial T}
\right)_{E}dE; \label{DT_int}\ee
where the pyroelectric coefficient is
\be \left(\frac{\partial P_3}{\partial T}
\right)_{E} = (e_{36}^{0}\varepsilon_T + \frac{2(\mu+12\mu'\eta^2)}{v}\eta_T); \ee
$V=vN_A/2$ is the molar volume.

\section{Numerical calculations}

To perform the numerical calculations we need to set the values of the following theory parameters
 \begin{enumerate}
 \item[-] The Slater energies
 $\varepsilon$, $w$, $w_{1}$ ;
 \item[-] the parameter of the long-range interactions $\nu_c$;
 \item[-] the effective dipole moment $\mu$;
 \item[-] the correction to the effective dipole moment due to proton ordering $\mu'$;
 \item[-] the deformation potentials  $\psi_{6}$,
 $\delta_{s}$, $\delta_{a}$,
$\delta_{1}$;
 \item[-] the ``seed''  dielectric susceptibility
 $\chi_{33}^{\varepsilon 0}$;
 \item[-] the ``seed'' elastic constant $c_{66}^{E0}$;
 \item[-] the ``seed'' piezoelectric coefficient
 $e_{36}^0$.
 \end{enumerate}

They are chosen, obviously, by fitting the theoretical thermodynamic characteristics to the experimental data, as described in  \cite{JPS1701}.

The energy $w_{1}$ of two proton configurations with four or zero
protons near the given oxygen tetrahedron should be much higher than
$\varepsilon$ and $w$. Therefore we take $w_{1} = \infty$ $(d=0)$.

The optimum sets of the model parameters are given in Table~1. $T_c^0$ is phase transition temperature at zero field.
\renewcommand{\arraystretch}{1.0}
\renewcommand{\tabcolsep}{3.0pt}
\begin{table}[!h]
\caption{The optimum sets of the model parameters for K(H$_{1 - x}$D$_{x})_{2}$PO$_{4}$.}
\begin{center}
\begin{tabular}{|c|c|c|c|c|c|c|c|c|c|}
\hline $x$ & $T_c^0$ & $\frac{\varepsilon}{k_B}$ & $\frac{w}{k_B}$ & $\frac{\nu_c}{k_B}$ & $\mu$ & $\mu'$ & $\chi_{33}^0$ \\
 & (K) & (K) & (K) & (K) & ($10^{-30}$~C$\cdot $m) & ($10^{-30}$~C$\cdot $m) & \\
%\hline   0.00 & 122.5 & 34.00 & 305.0 & 36.23 & 1.95*10^{-18} &  -0.13 &  0.73 \\
%\hline   0.00 & 122.22 & 68.00 & 605.0 & 8.19 & 5.07 &  -0.143 &  0.8 \\
\hline   0.00 & 122.22 & 56.00 & 430.0 & 17.55 & 5.6 &  -0.217 &  0.75 \\
   0.84 & 208.00 & 83.68 & 713.5 & 38.73 & 6.8 &  -0.217 &  0.41 \\
   0.88 & 211.00 & 85.00 & 727.0 & 39.17 & 6.8 &  -0.217 &  0.39 \\
   0.89 & 211.73 & 85.33 & 730.4 & 39.26 & 6.8 &  -0.217 &  0.39 \\
 \hline
\end{tabular}
\end{center}
\begin{center}
\begin{tabular}{|c|c|c|c|c|c|c|c|c|c|}
\hline $x$ & $\frac{\psi_{6}}{k_B}$ & $\frac{\delta_{s}}{k_B}$ & $\frac{\delta_{a}}{k_B}$ & $\frac{\delta_{1}}{k_B}$ & $c_{66}^{E0}$ & $e_{36}^0$ \\
 & (K) & (K) & (K) & (K) & ($10^{9}$~N/m$^2$) & (C/m$^2$) \\
\hline   0.00 &-150.00 & 82.00 &-500.00 &-400.00 &  7.00 &0.0033 \\
   0.84 &-140.45 & 51.45 &-977.27 &-400.00 &  6.43 &0.0033 \\
   0.88 &-140.00 & 50.00 &-1000.00 &-400.00 &  6.40 &0.0033 \\
   0.89 &-139.89 & 48.64 &-1005.68 &-400.00 &  6.39 &0.0033 \\
 \hline
\end{tabular}
\end{center}
\end{table}
\renewcommand{\arraystretch}{1}
\renewcommand{\tabcolsep}{1pt}

The primitive cell volume is taken to be  $v=0.195\cdot 10^{-21}$
cm$^3$ for all compositions. The values of the lattice specific
heat parameters of are $C_0=60$~J/(mol~K),
$C_1=0.32$~J/(mol~K$^2$) for $x=0$ and $C_0=93$~J/(mol~K),
$C_1=0.32$~J/(mol~K$^2$) for $x=0.86$ and 0.89.

\begin{figure}[!h]
\begin{center}
 \includegraphics[scale=0.65]{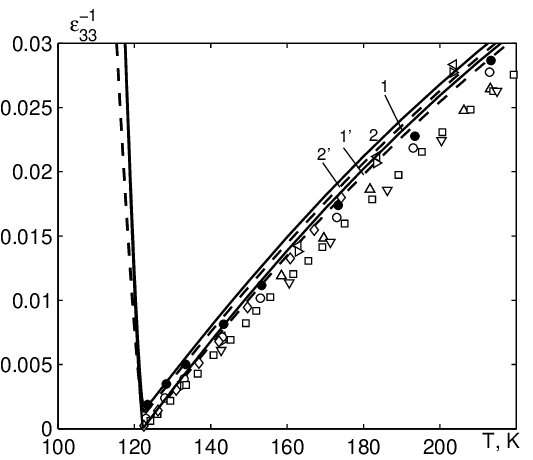}~~~ \includegraphics[scale=0.65]{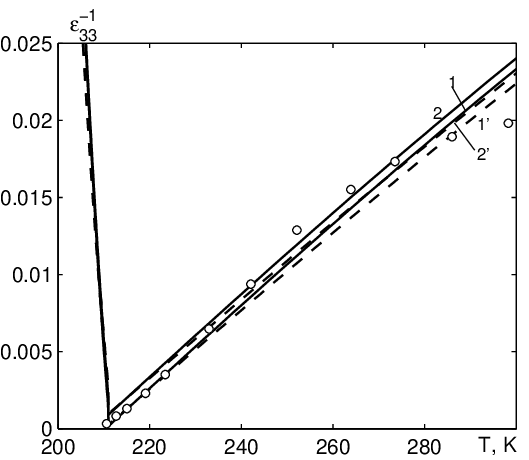}
\end{center}
\caption[]{The temperature dependence of the inverse static dielectric permittivities
of free  $(\varepsilon^{\sigma}_{33})^{-1}$ and clamped $(\varepsilon^{\varepsilon}_{33})^{-1}$  K(H$_{1 - x}$D$_{x})_{2}$PO$_{4}$ crystals
at $x=0.0$. Symbols are experimental data taken from  $\circ$, $\bullet$ -- \cite{138x},
$\square$ -- \cite{383x}, $\lozenge$ -- \cite{473x},
$\triangleright$ -- \cite{379x}, $\triangleleft$ -- \cite{369x},
$\triangledown$ -- \cite{375x}, $\vartriangle$\cite{481x}. Solid lines: the present theory; dashed
 lines: the theoretical results of  \cite{JPS1701} for $(\varepsilon^{\sigma}_{33})^{-1}$ (1') and $(\varepsilon^{\varepsilon}_{33})^{-1}$ (2').} \label{e33m1}
\caption[]{The same for  $x=0.88$. Symbols are experimental data taken from  $\circ$ -- \cite{371x}.} \label{e33m1_x88}
\end{figure}

\begin{figure}[!h]
\begin{center}
 \includegraphics[scale=0.65]{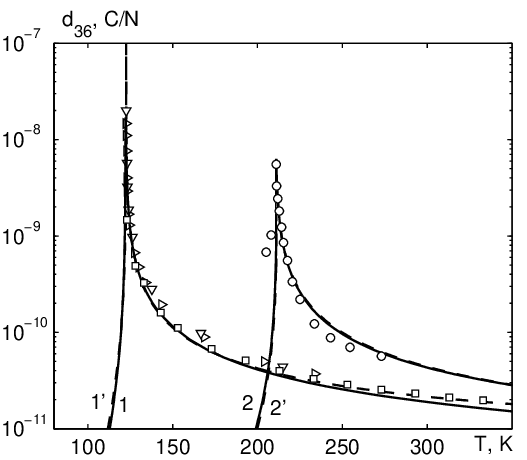}
\end{center}
\caption[]{The temperature dependence of the piezoelectric coefficient $d_{36}$ of K(H$_{1 - x}$D$_{x})_{2}$PO$_{4}$ at $x=0.0$ -- 1, 1', $\square$ \cite{138x}, $\triangledown$ \cite{bantle43},
$\triangleright$, \cite{von43};   at $x=0.88$ -- 2, 2', $\circ$ \cite{371x}. Dashed lines: the theoretical results of \cite{JPS1701}.} \label{d36_x88}
\end{figure}

\begin{figure}[!h]
\begin{center}
 \includegraphics[scale=0.65]{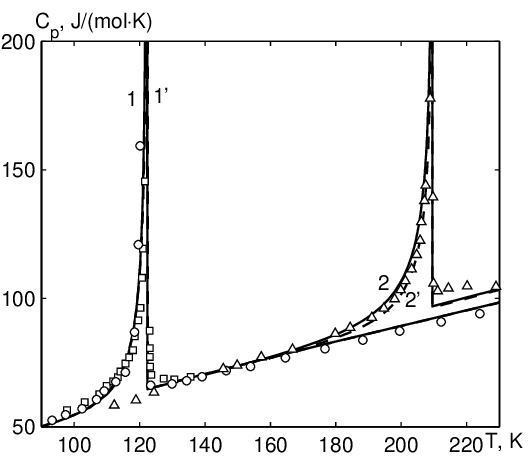}
\end{center}
\caption[]{The temperature dependence of the molar specific heat of  K(H$_{1 - x}$D$_{x})_{2}$PO$_{4}$ at $x=0.0$ -- $\circ$ \cite{Stephenson1397}, $\square$ \cite{380x}; at $x=0.86$ -- $\bigtriangleup$ \cite{380x}. Dashed lines: the theoretical results of \cite{JPS1701}.} \label{C_x86}
\end{figure}

When the dependence of the effective dipole moment on the order parameter is taken into account, the agreement between the theory and experiment
for most of the calculated thermodynamic characteristics of K(H$_{1 - x}$D$_{x})_{2}$PO$_{4}$ crystals in absence of the external electric field
is neither improved nor worsened. Thus, the calculated temperature dependences of the inverse static dielectric permittivities of free  $(\varepsilon^{\sigma}_{33})^{-1}$ and clamped $(\varepsilon^{\varepsilon}_{33})^{-1}$  crystals
 (figs.~\ref{e33m1}, \ref{e33m1_x88}), piezoelectric coefficient $d_{36}$  (fig.~\ref{d36_x88}), and molar specific heat (fig.~\ref{C_x86}) are close
 to the previous theoretical curves \cite{JPS1701}.

\begin{figure}[!h]
\begin{center}
 \includegraphics[scale=0.62]{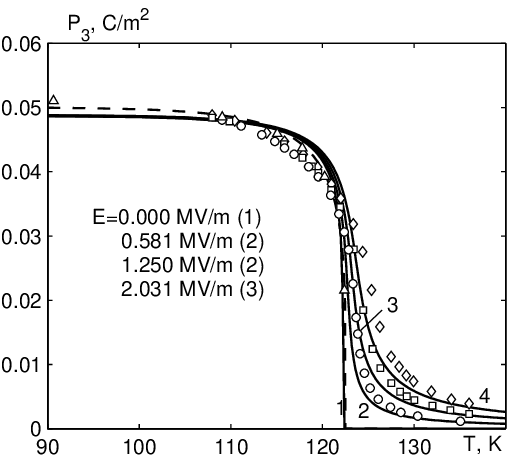}
\end{center}
\caption[]{The temperature dependence of polarization of
K(H$_{1 - x}$D$_{x})_{2}$PO$_{4}$ at $x=0$ and at different  $E_3$(MV/m): 0.0 -- 1,
$\vartriangle$ \cite{363x}; 0.581 -- 2, $\circ$\cite{369x};
1.250 -- 3, $\square$\cite{369x}; 2.031 -- 4,
$\lozenge$\cite{369x}. Symbols are experimental points; solid lines: the present theory; dashed
 lines: the theoretical results of  \cite{JPS1701}.} \label{Ps}
\end{figure}

\begin{figure}[!h]
\begin{center}
 \includegraphics[scale=0.625]{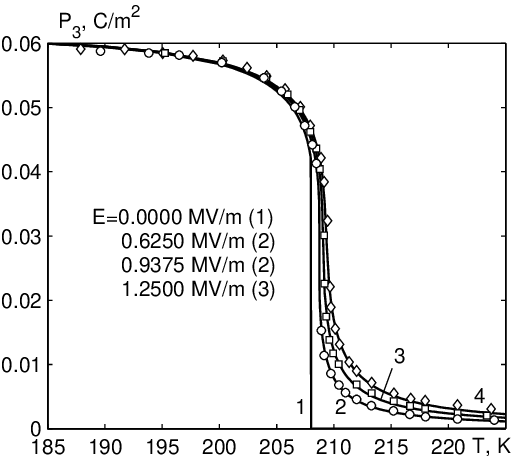}
\end{center}
\caption[]{The temperature dependence of polarization of
K(H$_{1 - x}$D$_{x})_{2}$PO$_{4}$ at $x=0.84$ and at different  $E_3$(MV/m): 0.0 -- 1; 0.625 -- 2, $\circ$;
0.9375 -- 3, $\square$; 1.25 -- 4, $\lozenge$. Symbols are experimental points; lines: the present theory.} \label{Ps_x84}
\end{figure}

\begin{figure}[!h]
\begin{center}
 \includegraphics[scale=0.62]{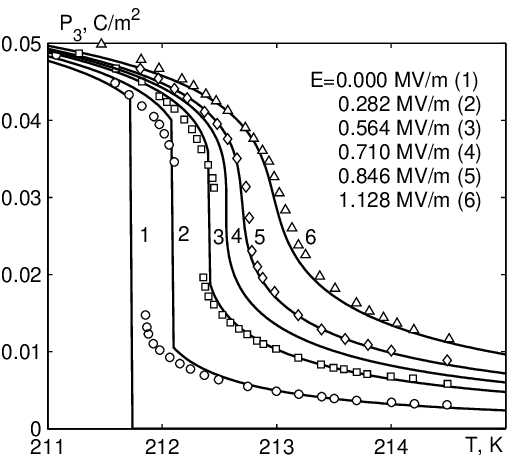}
\end{center}
\caption[]{The temperature dependence of polarization of
K(H$_{1 - x}$D$_{x})_{2}$PO$_{4}$ at $x=0.89$  and at different  $E_3$ (MV/m): 0.0 -- 1; 0.282 -- 2, $\circ$;
0.564 -- 3, $\square$; 0.71 -- 4; 0.846 -- 5, $\lozenge$; 1.128 -- 6, $\vartriangle$. Symbols are experimental points taken from \cite{Sidnenko978}; lines: the present theory.} \label{Ps_x89}
\end{figure}

However, the present model allows us to describe more consistently the smearing of the first order phase
in high electric fields. In figs.~\ref{Ps}, \ref{Ps_x84}, and \ref{Ps_x89} we plotted the temperature
variation of the polarization of  K(H$_{1 - x}$D$_{x})_{2}$PO$_{4}$ in different fields.

The agreement with experiment is better at  $x=0.84$ and 0.89 than
at $x=0$. We believe this is due to proton tunnelling, essential
in non-deuterated samples, which is not included in our model. The
field $E_{3}$, which in these crystals is the field conjugate to
the order parameter, induces non-zero polarization $P_{3}$ above
the transition point. Polarization has a jump at  $T_c$,
indicating the first order phase transition. With increasing
field, the polarization jump decreases, whereas the transition
temperature $T_c$ increases almost linearly. The corresponding
$\partial T_c/\partial E_3$ slopes are 0.192 and 0.115~K cm/kV for
$x=0$ and  $x=0.89$, respectively (c.f. 0.22 and 0.13~K cm/kV from
our earlier calculations \cite{Stasyuk2001} and experimental
$0.125$~K cm/kV of \cite{GladkiiSidnenko} for  $x=0.89$). At some
critical field $E^*$ the jump vanishes, and the transition smears
out. The calculated coordinates of the critical point are
 $E^*=125$~V/cm, $T^*_c$=122.244~K for $x=0$  and $7.1$~kV/cm, 212.55~K for $x=0.89$, which agrees well with the experiment \cite{Western,GladkiiSidnenko}.  It should be noted that in our previous
calculations \cite{JPS1701} it was impossible to obtain a correct
description of the polarization behavior in the fields above the
critical one, because of the necessity to use two different values
of the effective dipole moment $\mu$ in calculations.

\begin{figure}[!h]
\begin{center}
 \includegraphics[scale=0.63]{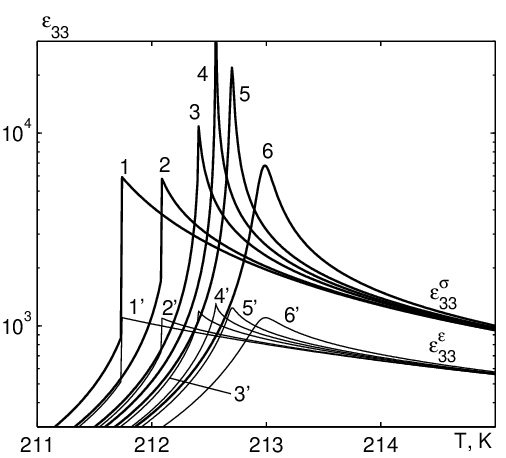}
\end{center}
\caption[]{The temperature dependence of the inverse static dielectric permittivities of free  $(\varepsilon^{\sigma}_{33})^{-1}$ (bold lines)
and clamped $(\varepsilon^{\varepsilon}_{33})^{-1}$ (thin lines)
 K(H$_{1 - x}$D$_{x})_{2}$PO$_{4}$ crystals for $x=0.89$ at different electric fields $E_3$ (MV/m): 0.0 -- 1, 1'; 0.282 -- 2, 2';
0.564 -- 3, 3'; 0.71 -- 4, 4'; 0.846 -- 5, 5'; 1.128 -- 6, 6'.} \label{e33_x89}
\end{figure}

\begin{figure}[!h]
\begin{center}
 \includegraphics[scale=0.63]{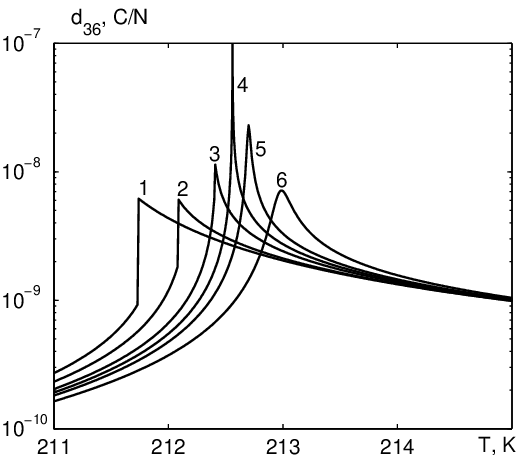}
\end{center}
\caption[]{The temperature dependence  of the piezoelectric coefficient
 $d_{36}$ of K(H$_{1 - x}$D$_{x})_{2}$PO$_{4}$ for $x=0.89$ at different electric fields $E_3$ (MV/m): 0.0 -- 1, 1'; 0.282 -- 2, 2'; 0.564 -- 3, 3'; 0.71 -- 4, 4'; 0.846 -- 5, 5'; 1.128 -- 6, 6'.} \label{d36_x89}
\end{figure}

\begin{figure}[!h]
\begin{center}
 \includegraphics[scale=0.62]{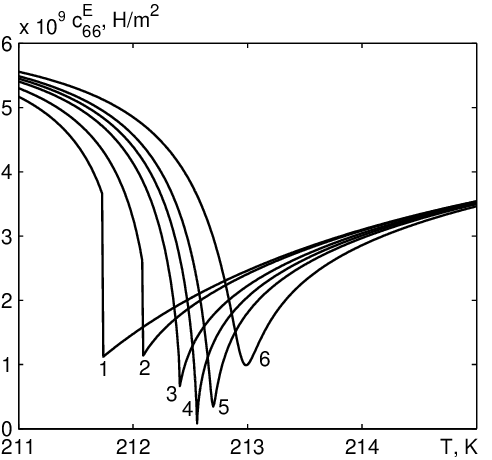}
\end{center}
\caption[]{The temperature dependence  of the elastic constant $c_{66}^E$ of K(H$_{1 - x}$D$_{x})_{2}$PO$_{4}$ for $x=0.89$ at different electric fields $E_3$ (MV/m): 0.0 -- 1, 1'; 0.282 -- 2, 2'; 0.564 -- 3, 3'; 0.71 -- 4, 4'; 0.846 -- 5, 5'; 1.128 -- 6, 6'.} \label{c66e_x89}
\end{figure}

Smearing of the phase transition is observed also in the temperature dependences of the dielectric permittivity $\varepsilon_{33}$ (fig.~\ref{e33_x89}), piezoelectric coefficient $d_{36}$ (fig.~\ref{d36_x89}), and elastic constant $c_{66}^E$ (fig.~\ref{c66e_x89}).

\begin{figure}[!h]
\begin{center}
 \includegraphics[scale=0.63]{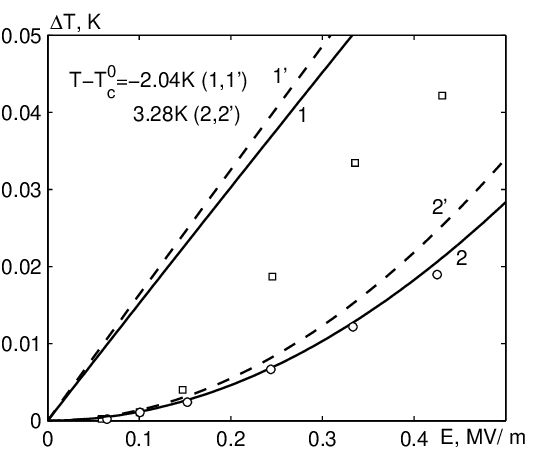}~~~~\includegraphics[scale=0.63]{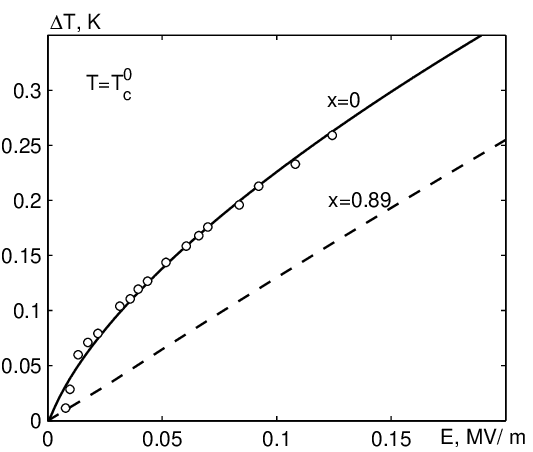}
\end{center}
\caption[]{The field dependences of the electrocaloric temperature change of  K(H$_{1 - x}$D$_{x})_{2}$PO$_{4}$ for $x=0.0$ (solid lines) in the ferroelectric phase at $T-T_c^0=-2.04$~K -- 1, $\square$ \cite{363x} and in the paraelectric phase at $T-T_c^0=3.28$~K -- 2, $\circ$ \cite{363x};
for $x=0.89$ (dashed lines)  $T-T_c^0=-2.04$~K -- 1' and $T-T_c^0$=3.28K -- 2'.} \label{DT_x89}
\caption[]{The field dependence of the electrocaloric temperature change of  K(H$_{1 - x}$D$_{x})_{2}$PO$_{4}$ at $T=T_c^0$ for $x=0.0$ (solid line, $\circ$ \cite{Shimshoni1969}) and $x=0.89$ (dashed line).} \label{DT_Tc_x89}
\end{figure}

\begin{figure}[!h]
\begin{center}
 \includegraphics[scale=0.63]{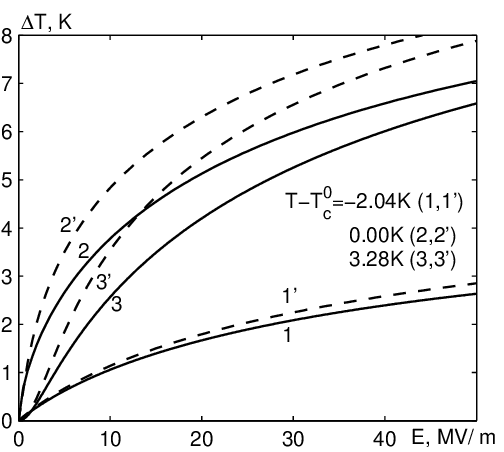}
\end{center}
\caption[]{The field dependence of the electrocaloric temperature
change of K(H$_{1 - x}$D$_{x})_{2}$PO$_{4}$ for $x=0.0$ (solid
lines) and  $x=0.89$ (dashed lines)  at $T-T_c^0=-2.04$~K -- 1,
$1'$; $T=T_c^0$ -- 2, $2'$; $T-T_c^0=3.2$~K -- 3, $3'$ for very
high fields.} \label{DT_x89_500kVcm}
\end{figure}

The calculated changes of temperature $\Delta T$ of the  KDP crystals with the adiabatically applied electric field is showin in figs.~\ref{DT_x89}, \ref{DT_Tc_x89}, and \ref{DT_x89_500kVcm}.
%%
%\begin{figure}[!h]
%\begin{center}
% \includegraphics[scale=0.63]{DT_x89.eps}
%\end{center}
%\caption[]{The field dependences of the electrocaloric temperature change of  K(H$_{1 - x}$D$_{x})_{2}$PO$_{4}$ for $x=0.0$ (solid lines) in the ferroelectric phase at $T-T_c^0=-2.04$~K -- 1, $\square$ \cite{363x} and in the paraelectric phase at $T-T_c^0=3.28$~K -- 2, $\circ$ \cite{363x};
%for $x=0.89$ (dashed lines)  $T-T_c^0=-2.04$~K -- 1' and $T-T_c^0$=3.28K -- 2'.} \label{DT_x89}
%\end{figure}
%
%\begin{figure}[!h]
%\begin{center}
% \includegraphics[scale=0.63]{DT_Tc_x89.eps}
%\end{center}
%\caption[]{The field dependence of the electrocaloric temperature change of  K(H$_{1 - x}$D$_{x})_{2}$PO$_{4}$ at $T=T_c^0$ for $x=0.0$ (solid line, $\circ$ \cite{Shimshoni1969}) and $x=0.89$ (dashed line).} \label{DT_Tc_x89}
%\end{figure}
%
As one can see, at small fields (fig.~\ref{DT_x89}) the calculated
electrocaloric temperature change is a linear function of the
field in the ferroelectric  (curves 1, $1'$) and a quadratic
function in the paraelectric phase (curves 2, $2'$). The
experimental behavior in the ferroelectric phase is not linear at
$E<3$~kV/cm because of the domains. The experimental data of
\cite{Shimshoni1969} (fig.~\ref{DT_Tc_x89}) were obtained at
$T=121$~K, which was very close to the transition temperature of
the sample used in the measurements.
 The domains, which polarization is  oriented along the field, are heated, whereas the domains, polarized in the opposite direction are cooled. The disagreement between the theory and experiment for an undeuterated crystal in the ferroelectric phase can be also caused by tunneling, which is not taken into account in the present model.
In very high fields (fig.~\ref{DT_x89_500kVcm}) the calculated
electrocaloric temperature change in the paraelectric phase are
larger than in the ferroelectric phase. The obtained curves
deviate from linear and quadratic behavior and reach saturation at
$E\gg50$~MV/m. To create fields that high in macroscopic single
crystals is obviously practically impossible, because of the
dielectric breakdown. However, experimental data for $\Delta T$
are not available even for moderate fields above 0.5~MV/m.

\begin{figure}[!h]
\begin{center}
 \includegraphics[scale=0.7]{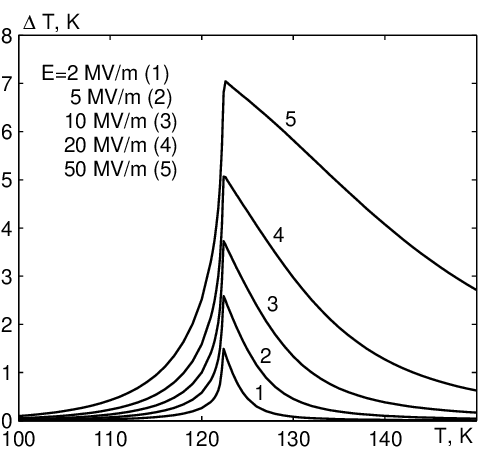}~~~\includegraphics[scale=0.7]{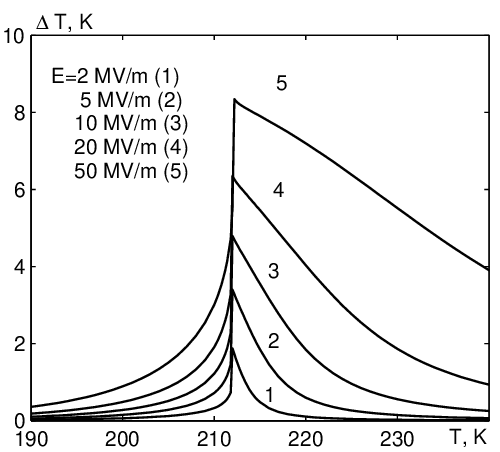}
\end{center}
\caption[]{The temperature dependence of the electrocaloric temperature change of K(H$_{1 - x}$D$_{x})_{2}$PO$_{4}$ for $x=0.0$ (left) and $x=0.89$ (right) in different fields.} \label{DT_T}
\end{figure}

As one can see from the temperature dependence of  $\Delta T$
(fig.~\ref{DT_T}), the calculated electrocaloric  temperature
change is the largest in the paraelectric phase close to $T_c$ and
can exceed  6~K.
The electrocaloric effect in K(H$_{1 - x}$D$_{x})_{2}$PO$_{4}$ at $x=0.89$ is larger than at $x=0.0$,
because with increasing deuteration the first order character of the phase transitions becomes more pronounced.

\begin{figure}[!h]
\begin{center}
 \includegraphics[scale=0.7]{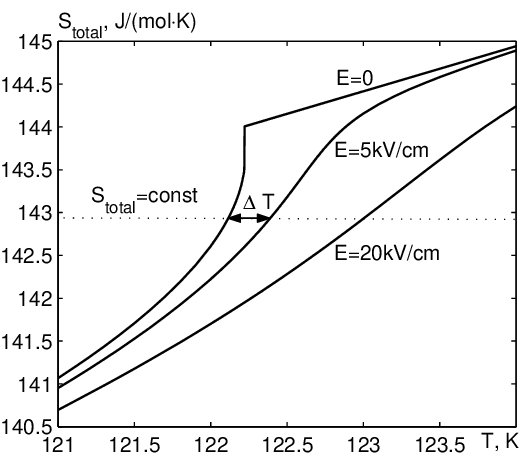}
\end{center}
\caption[]{The temperature dependence of molar entropy of  KDP at different fields.} \label{S_fig}
\end{figure}

We can also find $\Delta T$ using Eq.~(\ref{DT_S}), that is, as illustrated in fig.~\ref{S_fig}.
The values of  $\Delta T$ calculated using Eqs.~(\ref{DT_int}) and (\ref{DT_S}) coincide.

 \section{Conclusions}

Taking into account the dependence of the effective dipole moment on the order parameter allows us to correctly describe
smearing of the ferroelectric phase transition in high electric field as well as the electrocaloric effect in KDP crystals.
The theory predicts the values of the electrocaloric temperature  change above 5~K in very high fields. This fact could make
 the  KDP crystals a promising material for electrocaloric refridgerators.
Additional experimental measurements of  $\Delta T$ in fields
above 0.5~MV/m are necessary.

%\begin{center}
%\textbf{LONGITUDINAL DIELECTRIC, PIEZOELECTRIC, ELASTIC AND
%THERMAL PROPERTIES OF KH$_2$PO$_4$ TYPE FERROELECTRICS}
%\end{center}
%
%\begin{center}
%R.R.Levitsky$^a$, I.R.Zachek$^b$, A.S.Vdovych$^a$, A.P.Moina$^a$
%\end{center}
%
%\begin{center}
%$^a$\textit{Institute for Condensed Matter Physics \\
%of the National Academy of Sciences of Ukraine, \\ 1 Svientsitskii
%Str., 79011, Lviv, Ukraine \\
%$^b$National University ''Lvivska Politechnika''\\
%12 S. Bandera Str., 79013, Lviv, Ukraine}
%\end{center}
%
%
%
%Within the framework of modified model of proton ordering of
%KH$_{2}$PO$_{4}$ type ferroelectrics with taking into account
%linear on strain $\varepsilon _{6}$ contribution into the energy
%of proton system, but without taking into account tunneling within
%the four particle cluster approximation corresponding
%thermodynamic potentials are calculated. Using the  corresponding
%equations of state spontaneous polarization are calculated
%longitudinal dielectric permittivity of a mechanically clamped and
%mechanically free crystals, their piezoelectric characteristics,
%elastic constants and molar capacity. At the proper set of
%parameters good quantitative description of the available
%experimental data for the K(H$_{1 - x}$D$_{x})_{2}$PO$_{4}$ type
%ferroelectrics is obtained.
%
%Keywords: ferroelectrics, cluster approximation, dielectric
%permittivity, piezoelectric modulus, elastic constant.
%
%
%PACS numbers: 77.80.жe, 77.84.жs, 77.84.Fa, 77.65.Bn

\end{document}